\documentclass[prl,floatfix,superscriptaddress,twocolumn,aps,showpacs]{revtex4}

\usepackage{amssymb} 
\usepackage{graphicx} 
\usepackage{longtable}
\usepackage{amsmath} 
\usepackage{amsfonts} 
\usepackage{color}
\usepackage{times} 
\usepackage{stmaryrd} 
\usepackage[latin1]{inputenc}

\begin{document}

\title{Landau-Zener transition in a continuously measured single-molecule spin transistor}  

\author{F.~Troiani} 
\affiliation{Centro S3, Istituto Nanoscienze - CNR, via G. Campi 213/A, I-41125 Modena, Italy} 
\author{C.~Godfrin} 
\affiliation{Institut L. N\'eel, CNRS, Av des Martyrs 25,  F-38000 Grenoble, France} \author{S.~Thiele} 
\affiliation{Institut L. N\'eel, CNRS, Av des Martyrs 25,  F-38000 Grenoble, France} 
\author{F.~Balestro} 
\affiliation{Institut L. N\'eel, CNRS, Av des Martyrs 25,  F-38000 Grenoble, France} 
\author{W.~Wernsdorfer} 
\affiliation{Institut L. N\'eel, CNRS, Av des Martyrs 25,  F-38000 Grenoble, France} 
\affiliation{Institute of Nanotechnology, Karlsruhe Institute of Technology (KIT),  D-76344 Eggenstein Leopoldshafen, Germany} 
\author{S.~Klyatskaya} 
\affiliation{Institute of Nanotechnology, Karlsruhe Institute of Technology (KIT),  D-76344 Eggenstein Leopoldshafen, Germany} 
\author{M.~Ruben} 
\affiliation{Institute of Nanotechnology, Karlsruhe Institute of Technology (KIT),  D-76344 Eggenstein Leopoldshafen, Germany} 
\author{M.~Affronte} \affiliation{Centro S3, Istituto Nanoscienze - CNR, via G. Campi 213/A,  I-41125 Modena, Italy} 
\affiliation{Dipartimento di Scienze Fisiche, Informatiche e Matematiche, Universit\`a di Modena e Reggio Emilia, via G. Campi 213/a, I-41125 Modena, Italy}  

\begin{abstract} 
We monitor the Landau-Zener dynamics of a single-ion magnet in a spin-transistor geometry. For increasing field-sweep rates, the spin reversal probability shows increasing deviations from that of a closed system. In the low-conductance limit, such deviations are shown to result from a dephasing process. In particular, the observed behaviors are succesfully simulated by means of an adiabatic master equation, with time averaged dephasing (Lindblad) operators. The time average is tentatively interpeted in terms of the finite time resolution of the continuous measurement. 
\end{abstract}

\date{\today }


\maketitle

{\it Introduction ---} The dynamics of a quantum system driven through an avoided level crossing represents a relevant problem in many physical contexts. In the simplest case, known as the Landau-Zener problem, the dynamics involves only two states, coupled by a constant tunneling term, and separated in energy by a gap that depends linearly on time. This problem was independently solved by different authors, who provided an analytical expression for the probability that the system eventually undergoes a spin reversal \cite{Landau32,Zener32,Majorana32,Stueckelberg32}. However, these results apply to the ideal case of an isolated quantum system. In realistic conditions, coupling to the environment tends to induce decoherence, both through elastic and inelastic processes. In fact, decoherence can affect the Landau-Zener dynamics in substantially different ways, depending on the environment composition (e.g., harmonic oscillators or spins), temperature, and on the presence or absence of memory effects \cite{Kayanuma84,Ao89,Saito02,Sinitsyn03,Saito07,Nalbach09}. Departures from a unitary evolution can also be induced by a measurement process, which
presents deep conceptual and formal connections with decoherence \cite
{Breuer,Braginsky,Mensky}. In particular, a continuous measurement of the system tends to destroy the phase coherence between different eigenstates of the observable. Continuous measurements of single quantum objects have been investigated by electrical and optical means in mesoscopic \cite{Gurvitz97,Bucks98,Korotkov99,HsiSheng01} and atomic systems \cite{Haikka14}, respectively. 
Whether decoherence is induced by the coupling to a quantum environment or to a measuring apparatus, its effective character qualitatively depends on the interplay between such coupling and the interactions within the system \cite{Paz99}, which can be explored and controlled by means of an external drive \cite{Katz}.

\begin{figure}
\begin{center}
\includegraphics[width=0.45\textwidth]{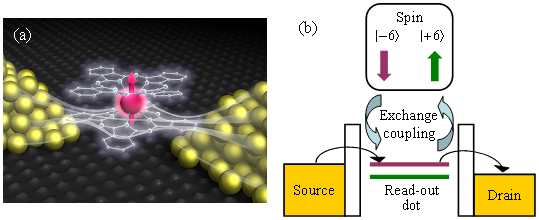}
\includegraphics[width=0.475\textwidth]{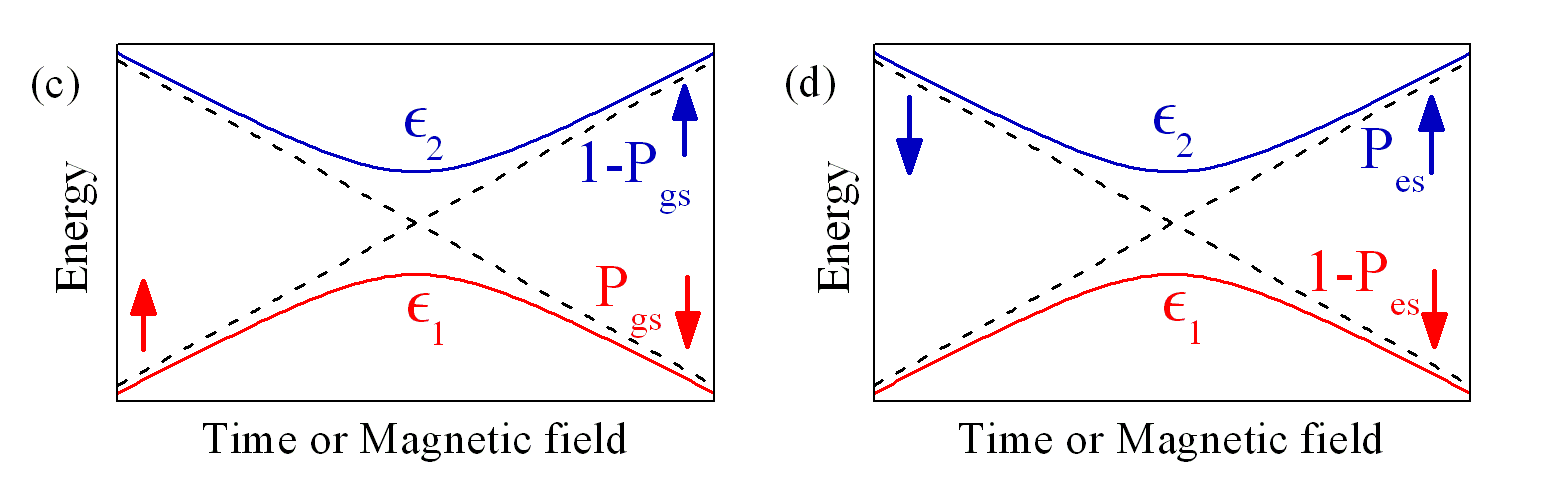}
\caption{\label{fig1}
(a) Artistic view of the molecular spin transistor with the TbPc$_2$ molecule embedded between the gold electrodes. 
(b) Schematics of the molecular system: the phthalocyanine acts as a read-out quantum dot, where the spins of the localized electrons are exchange coupled to the total angular momentum ($J=6$) of the Tb$^{3+}$ ion, whose $M_{J} = \pm6 $ states define an effective two-level system. 
(c) The system is prepared in the initial ground state 
$|\!\uparrow\rangle$, 
evolves under the effect of a magnetic field $B_z$ that depends linearly on time, and ends up in the final ground state $|\!\downarrow\rangle$ with probability $P_{gs}$. 
(d) If the system is prepared in the initial excited state, the spin reversal leads to the final excited state with probability $P_{es}$.}
\end{center}
\end{figure}

Here we experimentally and theoretically investigate the Landau-Zener dynamics of a single-ion magnet that is continuously measured by current within a molecular spin transistor geometry. The observed dependence of the spin-reversal probability on the  field-sweep rate presents clear deviations from the Landau-Zener formula, and thus from the behavior of an isolated quantum system. The weak dependence of the spin-reversal probability on the initial (ground or excited) spin state, indicates that such deviations are essentially due to dephasing, rather than relaxation or incoherent excitation. In order to account for the experimental results, we simulate the spin dynamics through a phenomenological master equation. The simulations suggest that the form of dephasing affecting the spin depends on the time scale of the Landau-Zener process so that the decoherence process becomes less effective in the limit of an adiabatic time evolution. This also explains why, rather counterintuitively, deviations from a coherent behavior are more significant for high field-sweep rates than for slow spin dynamics. 

{\it Experiment ---}
The single-molecule spin transistor consists of a single-ion magnetic molecule (TbPc$_2$), which is trapped between two gold electrodes, obtained by electromigration \cite{Vincent12} [Fig.~\ref{fig1}(a)]. A large spin-orbit coupling, in combination with a strong ligand field interaction, yields a well isolated electron ground state doublet ($J \!=\! 6$, $M_{J} = \pm 6 $) of the Tb$^{3+}$ ion, with a uniaxial anisotropy axis perpendicular to the phthalocyanine plane. Non-axial terms in the ligand-field Hamiltonian couple the $ M_{J} \!=\! \pm 6 $ states, giving rise to a zero-field energy gap $\Delta$ in the $\mu$K range. In the following, we refer to the Tb ion as an effective two-level system and label its $ M_{J} \!=\! + 6 $ and $ M_{J} \!=\! - 6 $ states with $|\!\uparrow\rangle$ and $|\!\downarrow\rangle$, respectively. The electron-spin is driven through an avoided level crossing by a time-dependent magnetic field, applied along the direction of the uniaxial anisotropy. The field sweep eventually induces an electron spin reversal with a probability $P$, whose dependence on the system and driving parameters is the main object of the present investigation. Instead, we eliminate the dependence of the above dynamics on the $I=3/2$ nuclear spin of the Tb$^{3+}$ ion by averaging on the four values of $M_I$.

In order to measure $P$, we sweep the magnetic field back and forth between $\pm 80\,$mT ($10^3$ times for each sweeping rate) and record the frequency of the spin-flip events. Typical sweeping rates $dB/dt$ range between 1 to 100 mT/s, such that the time that the system takes to go through the anticrossing typically ranges between 1 and $100\,\mu$s. During the field sweep, the electron-spin dynamics is monitored by the current that flows between the metallic source and drain electrodes and through the molecule. In the neutral TbPc$_2$ molecule, the valence of the Tb$^{3+}$ ion is not perturbed, due to the large ionization energy required for this process \cite{zhu04}. The phatholocianine constitutes an ideal molecular quantum dot, where the electrons couple to the Tb magnetic moment through an exchange interaction [Fig.~\ref{fig1}(b)]. Such interaction results in a dependence of the conductance on the Tb spin, and specifically in a change of the conductance of about 4\% in the case of a spin reversal \cite{Vincent12}. Therefore, each electron tunneling through the molecule's read-out dot weakly probes the spin state, which is continuously measured by the cumulative effect of the many tunneling events occurring during each field sweep. Such monitoring of the spin state, averaged over the $10^3$ field sweeps, can be regarded as an unread measurement.

\begin{figure}
\begin{center}
\includegraphics[width=0.5\textwidth]{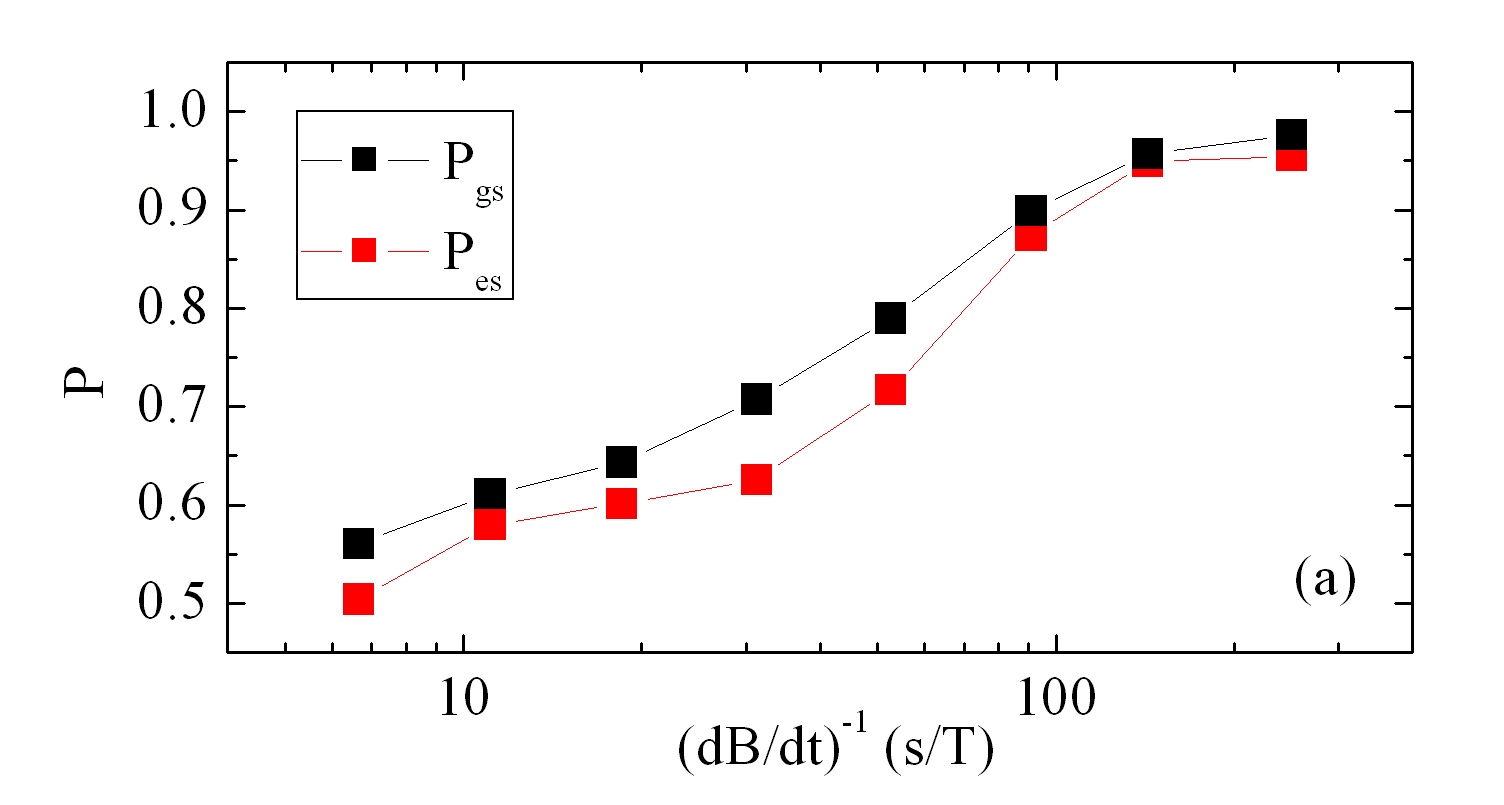}
\includegraphics[width=0.5\textwidth]{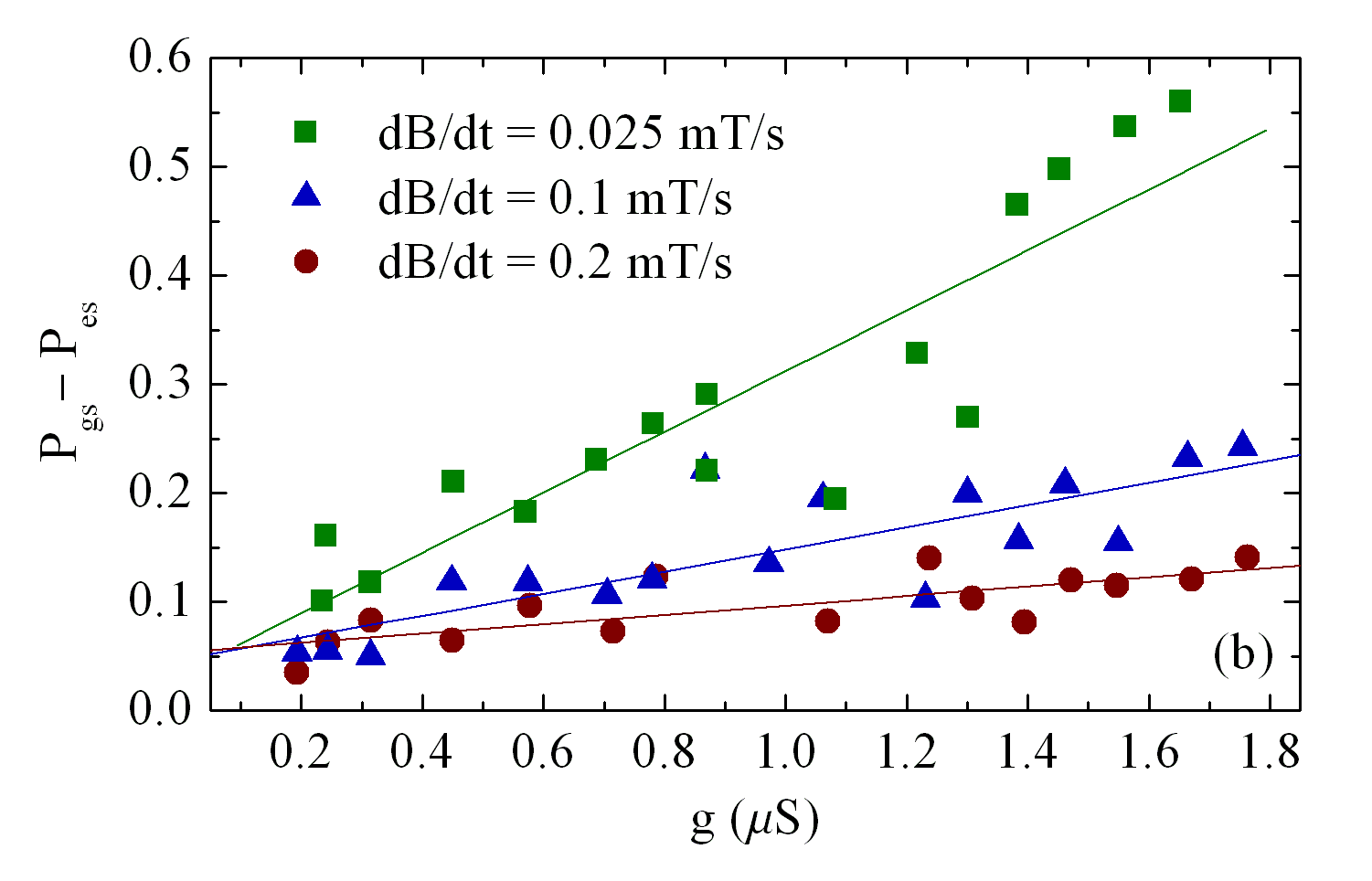}
\caption{\label{fig2}
(a) Measured values of the spin-reversal probabilities $P_{gs}$ (black squares) and $P_{es}$ (red), obtained after preparing the spin in the initial ground and exited states, respectively (the solid lines are drawn as a guide for the eye). This set of probabilities has been obtained with a conductance of $ g = 0.245\,\mu$S. 
(b) Difference between $P_{gs}$ and $P_{es}$ as a function of the conductance, for different values of the field-sweep rate $(dB/dt)$. The solid lines correspond to linear fits of the experimental results (symbols).
}
\end{center}
\end{figure}

We start by considering the spin-reversal probability corresponding to a transition from the initial to the final ground state ($P_{gs}$) as a function of the sweeping rate [Fig.~\ref{fig1}(c)]. The measurements are performed on two different devices and at very low temperature (30 mK). The observed dependence of $P_{gs}$ on the sweeping rate [Fig.~\ref{fig2}(a), black squares] significantly deviates from the Landau-Zener behavior. In particular, for large values of $dB/dt$ the probability approximately saturates at $0.5$, rather than at 0, as would be expected for a closed quantum system. Such deviation represents a strong indication that decoherence plays a role in the present dynamics. In particular, the increase of $P_{gs}$ for decreasing sweeping rates might be due to spin relaxation, which can in principle be induced by the coupling of the Tb spin with vibrations or with neighboring spins \cite{Wernsdorfer05}. In order to single out the role of such inelastic processes, we compare $P_{gs}$ with the probability $P_{es}$ of a transition from the initial to the final excited state [Fig.~\ref{fig1}(d)]. Here we find that the difference between $P_{gs}$ and $P_{es}$ is significant for relatively high current intensities, indicating the presence of an efficient spin relaxation mechanism [Fig.~\ref{fig2}(b)]. For smaller currents, however, the difference between the two probabilities vanishes and the system enters a regime where spin relaxation is ineffective. In fact, in such regime the dependence on the sweeping rate of $P_{es}$ follows quite closely that of $P_{gs}$ [Fig.~\ref{fig2}(a), red squares]. This allows us to exclude that, at least in the limit of vanishingly small currents flowing through the molecular dot, inelastic processes are responsible for the observed deviations from the Landau-Zener behavior. Such conclusion is further corroborated by the simulation of the spin dynamics in the presence of relaxation and incoherent excitation processes \cite{SM}. 

{\it Theory ---}
The spin dynamics results from the interplay between the time-dependent magnetic field and the constant tunneling term. Such interplay is described by the Hamiltonian:
\begin{equation}
\label{eq1}
H(t) 
= 
\frac{\alpha}{2} \left(t-\frac{T}{2}\right) \sigma_z + \frac{\Delta}{2} \sigma_x 
= 
\sum_{k=1}^2 \epsilon_k (t) | \epsilon_k(t) \rangle\langle \epsilon_k(t) | ,
\end{equation}
where the Pauli operators are expressed in the {\it diabatic basis} $\{ |\!\uparrow\rangle , |\!\downarrow\rangle \}$, while the time-dependent eigenstates of the Hamiltonian define the {\it adiabatic basis} $\{ |\epsilon_1(t)\rangle , |\epsilon_2(t)\rangle \}$. The parameters entering the above Hamiltonian are: the duration of the magnetic field sweep ($T$), the rate of variation of the Zeeman splitting ($\alpha = g_J \mu_B\, dB/dt$, with $g_J=18$ the g-factor of the effective two-level system), and the transverse coupling between the $M_J=+6$ and $M_J=-6$ states ($\Delta$). In the case of a closed system, the spin-reversal probability depends on these parameters through the Landau-Zener formula \cite{Landau32,Zener32,Majorana32,Stueckelberg32}:
\begin{equation}
\label{eq2}
P_{LZ} 
= 
1 - e^{-\frac{\pi}{2} \frac{\Delta^2}{\hbar\alpha}}
\equiv
1 - e^{-\frac{\pi}{2} \frac{\tau_{ac}}{\tau_\Delta}} .
\end{equation}
For the sake of the following discussion, we have introduced $\tau_\Delta \equiv \hbar / \Delta$ and $\tau_{ac} \equiv \Delta / \alpha$, which can be identified respectively with the characteristic time scale of the spin tunneling and with the time that the system takes to go through the level anticrossing. The spin reversal probability $P_{LZ}$ thus increases from 0 to 1 as the system passes from the diabatic ($\tau_{ac} \ll \tau_\Delta$) to the adiabatic regime ($\tau_{ac} \gg \tau_\Delta$). 

The coupling of the system to an environment can substantially modify the dependence of the spin reversal probability $P$ on the sweeping rate. In the case of a Markovian decoherence, the effect of such coupling can be simulated by means of a master equation in the Lindblad form \cite{Breuer}:
\begin{equation}
\label{eq3}
\dot{\rho} 
= 
\frac{i}{\hbar} [\rho , H] 
+ 
\sum_k
\left( 2 L_k \rho L_k^\dagger - L_k^\dagger L_k \rho - \rho L_k^\dagger L_k \right) ,
\end{equation}
where the Lindblad operators $L_k$ describe different forms of measurement or decoherence processes. The comparison between the measured probabilities $P_{gs}$ and $P_{es}$, as well as the simulations of inelastic processes (relaxation and incoherent excitation) \cite{SM}, shows that the dominant decoherence mechanism is here represented by dephasing, on which we focus in the following. In a two-level system, the loss of phase coherence between two states can be described by a Lindblad operator proportional to the difference between the projectors on such states. In the prototypical cases, hereafter labeled $a$ and $b$, dephasing takes place between the states that form either the diabatic or the adiabatic basis \cite{Novelli15}:
\begin{equation}
\label{eq4}
L_a
=
\frac{\sigma_z}{2\sqrt{\tau_d}}, \ 
L_b(t)
=
\frac{\eta}{2\sqrt{\tau_d}}
\left[ 
|\epsilon_1(t) \rangle\langle \epsilon_1(t)|
-
|\epsilon_2(t) \rangle\langle \epsilon_2(t)| 
\right] ,
\end{equation}
where $\eta = \langle\epsilon_1(t)|\sigma_z|\epsilon_1(t)\rangle$ and $\tau_d$ is the dephasing time. The case ($a$) corresponds to a loss of phase coherence between the diabatic states, which results from a system-environment coupling larger than the system self-Hamiltonian \cite{Paz99}. In the case ($b$), instead, dephasing affects the relative phase between the time-dependent eigenstates, as occurs if the self-Hamiltonian  represents the dominant term and its variation in time is slow enough to induce an adiabatic time evolution \cite{Albash12}. In the absence of a detailed knowledge of the physical environment experienced by the Tb spin, we cannot determine {\it a priori} whether the system falls into one of the above regimes or in some intermediate one. From a phenomenological perspective, however, we note that the dependence of the simulated spin-reversal probabilities on the sweeping rate obtained in the two prototypical cases ($P_a$ and $P_b$) clearly differs from the measured ones. In particular, $P_a$ saturates to 0.5 for small sweeping rates (i.e. for $ \tau_{ac} \gtrsim \tau_d $, not shown), whereas the measured probabilities $P_{gs}$ and $P_{es}$ saturate to 1. On the other hand, $P_b$ tends to 0 for high sweeping rates (and, in fact, hardly differs from the Landau-Zener probability, as shown below), where the measured probabilities tend to 0.5. 

\begin{figure}
\begin{center}
\includegraphics[width=0.45\textwidth]{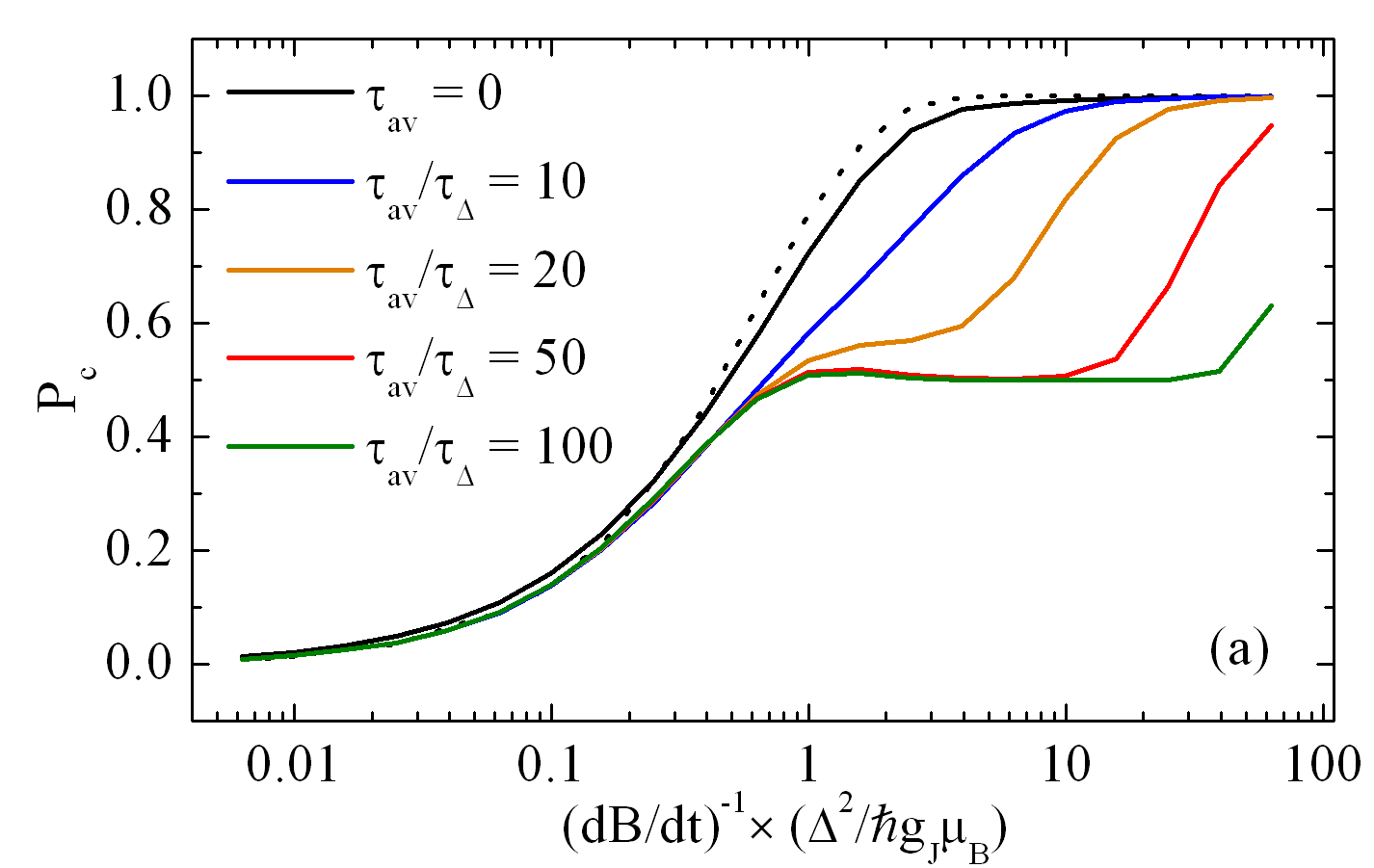}
\includegraphics[width=0.45\textwidth]{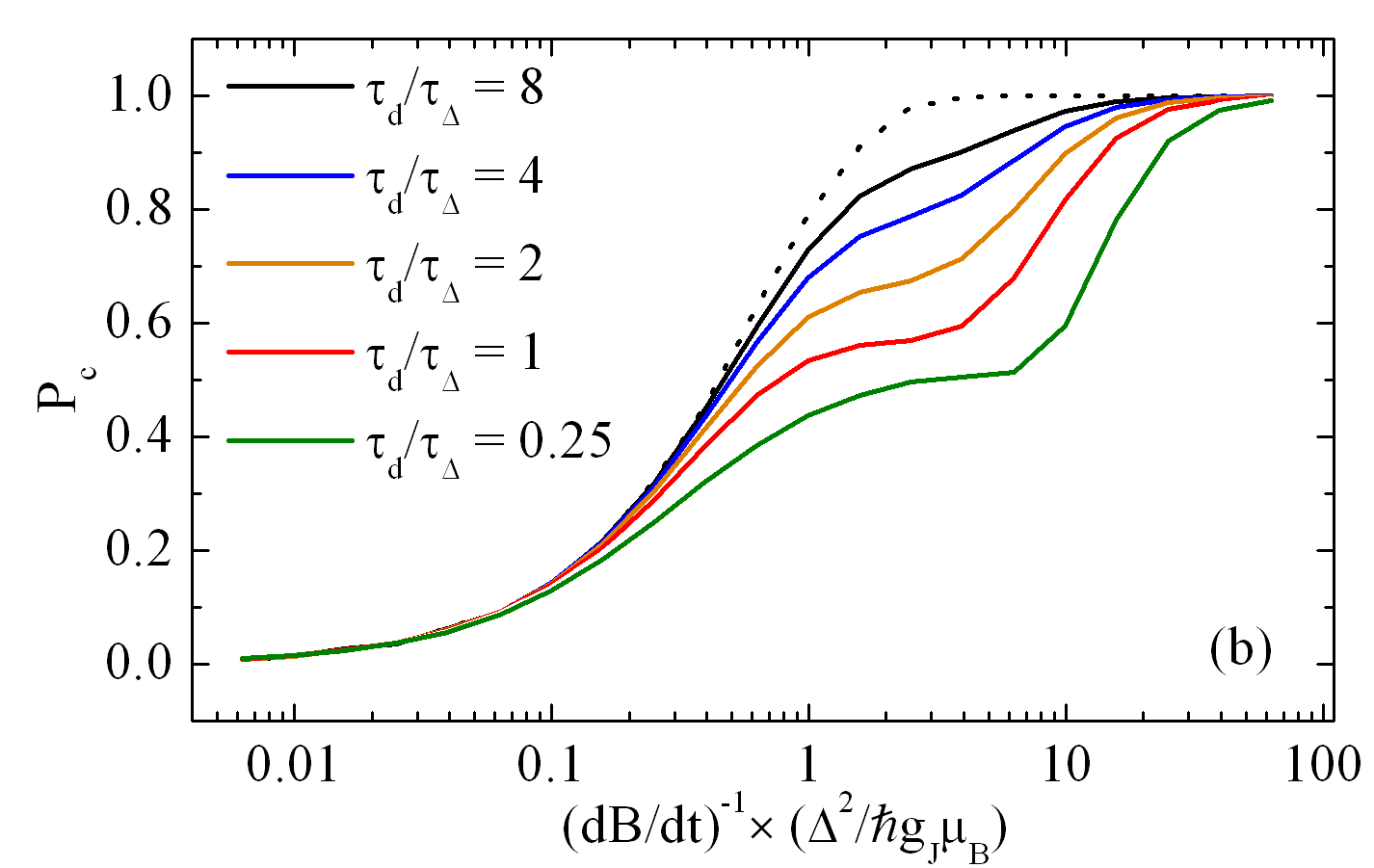}
\caption{\label{fig3}
(a) Computed spin-reversal probability $P_c$ as a function of the inverse field-sweep rate, for different values of the averaging time $\tau_{av}$, normalized to $\tau_\Delta$. For a given $\Delta$, the quantity reported in the horizontal axis can also be identified with the time that the spin takes to go through the anticrossing, being $ \tau_{ac} / \tau_\Delta = ( \Delta^2 / \hbar g_J \mu_B ) (dB/dt)^{-1}$. For all the solid curves, the dephasing time is $\tau_d=\tau_\Delta$, while the dotted curve corresponds to the coherent case ($\tau_d = \infty$). 
(b) Dependence of $P_c$ on the inverse sweeping rate for a fixed averaging time, $\tau_{av} = 20\,\tau_\Delta$, and for different values of the dephasing time $\tau_d$.
}
\end{center}
\end{figure}

In order to account for the observed behavior and to gain further insight into the decoherence process, we introduce a phenomenological master equation, where the Lindblad operator reflects the time dependence of the system eigenstates, as in case $(b)$, but with a finite time resolution. This is formalized by a time average over an interval of length $\tau_{av}$:
\begin{equation}
\label{eq5}
L_c(t; \tau_{av})
=
\frac{1}{\tau_{av}}
\int_{t-\tau_{av}/2}^{t+\tau_{av}/2}
\,
L_b(\tau)
\,
d\tau .
\end{equation}
The spin-reversal probability obtained by solving the master equation Eq. (\ref{eq3}) defined by the above Lindblad operator $L_c$ is labeled $P_c$. In the limiting cases where $\tau_{av} $ is much larger or much smaller than $\tau_{ac}$, $L_c$ coincides respectively with $L_a$ and $L_b$, such that one recovers the previously considered master equations. Therefore, by modifying the field sweeping rate one effectively changes the form of the system-environment interaction, formally represented by $L_c$, and the resulting decoherence process. Hereafter we show how this affects the dependence of $P_c$ on the field-sweep rate

\begin{figure}
\begin{center}
\includegraphics[width=0.475\textwidth]{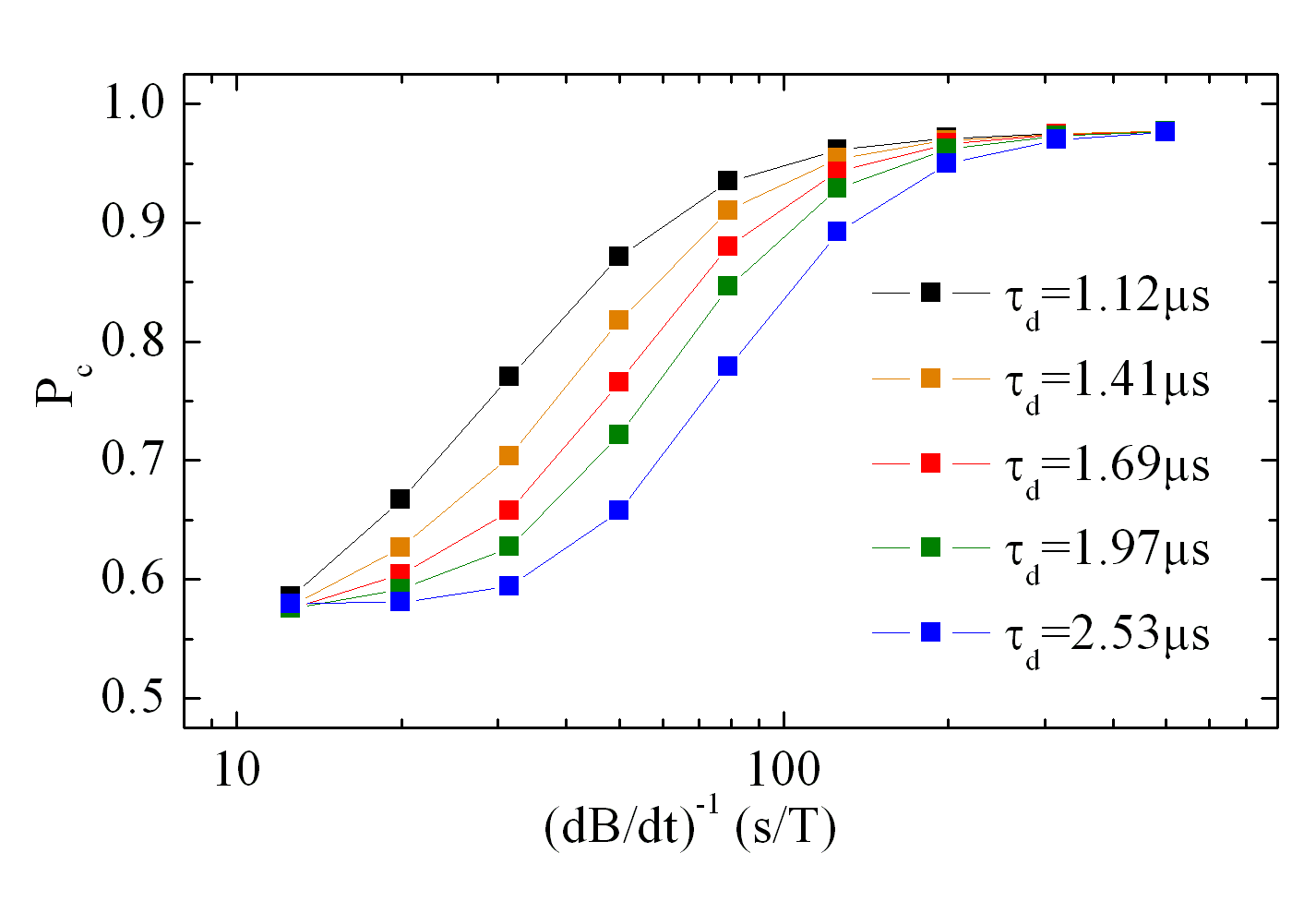}
\caption{\label{fig4}
Simulated values of the spin-reversal probability as a function of the inverse sweeping rate, for different values of the dephasing time $\tau_d$ and of the averaging time $\tau_{av} = 20 \tau_d$. The reported values of the inverse sweeping rate and of the dephasing time correspond to a zero-field gap $\Delta \simeq 3.4\,\mu$K or, equivalently, to a time scale $\tau_\Delta \simeq 2.25\,\mu$s.}
\end{center}
\end{figure}

The behavior of the spin-reversal probability $P_c$ can be essentially rationalized in terms of the relation between $\tau_{ac} = (\Delta / g_J\mu_B)(dB/dt)^{-1} $ and the time scales $\tau_{d}$ and $\tau_{av}$. We start by considering the dependence of $P_c$ on $\tau_{av}$ in the representative case where $\tau_d$ equals the tunneling time $\tau_\Delta$. For $\tau_{av}=0$ [Fig.~\ref{fig3}(a), solid black curve], $P_c$ coincides by definition with $P_b$ and hardly differs from the Landau-Zener probability $P_{LZ}$, corresponding to the coherent dynamics (dotted curve). For larger values of the averaging time, and specifically for $\tau_{av} \gtrsim \tau_\Delta$, the trend of $P_c$ changes qualitatively and a plateau at 0.5 appears, besides those at 0 and 1. We note that the rise of the spin-reversal probability from 0 to 0.5 and that from 0.5 and 1 have different physical origins. The former one results from the coherent part of the dynamics, and specifically from the fact that the system approaches the adiabatic regime. The latter rise, which resembles the one observed in the experimental curves [Fig.~\ref{fig2}(a)] and occurs at $\tau_{ac} \simeq \tau_{av} $, is instead due to the incoherent contribution. In particular, it can be traced back to the transition from a dephasing process between the diabatic states to one between the adiabatic states. The dependence of the spin reversal probability on the dephasing time $\tau_d$, for a given $\tau_{av}$, presents different features [Fig. \ref{fig3}(b)]. In fact, in first approximation, $\tau_d$ determines to which extent the spin-reversal probability is decreased with respect to the coherent case (dotted curve) in a given range of sweeping-rate values, which is determined by $\tau_{av}$.

The above results outline the general dependence of $P_c$ on the relevant time scales. Besides, they allow us to identify the ratios $\tau_d / \tau_\Delta$ and $\tau_{ac} / \tau_\Delta$ for which $P_c$ reproduces the observed functional dependence of the spin-reversal probability on the sweeping rate. The value of $\Delta$, and thus the absolute values of all the time scales, can be estimated by requiring that $P_c$ quantitatively agrees with $P_{gs}$ and $P_{es}$ for each given value of $(dB/dt)$. Such an agreement is found for a reasonable value of the zero-field splitting \cite{Ishikawa}, $\Delta \simeq 3.4\,\mu$eV, and leads to an estimate of the dephasing time of the order of a few $\mu$s (Fig. \ref{fig4}). This estimate is consistent with what expected for the environment-induced dephasing of a molecular spin at low temperatures \cite{Ghirri} and also corresponds to the expected time scale of a measurement-induced dephasing in the present experimental set up \cite{SM}.

In conclusion, our combined experimental and theoretical investigation provides clear evidence that a dephasing process affects the Landau-Zener dynamics of the molecular spin. 
The overall dependence of the spin-reversal probability on the field-sweep rate is reproduced by an adiabatic master equation, with time averaged Lindblad operators. As a result, the effective character of the dephasing process qualitatively depends on the time scale of the spin-reversal and decoherence is less effective for slow (adiabatic) spin manipulation. At a quantitative level, the comparison between experimental and theoretical results leads to an estimate of the system parameters (zero-field splitting and decoherence time) which is consistent with the expected values. Further investigation is needed in order to establish to which extent the observed decoherence is induced by the quantum environment or by the back action of the continuous measurement, which are expected to act on comparable time scales in the present device. In the case of a measurement-induced dephasing, the time average of the dephasing operators can account for the finite time resolution of the continuous measurement. 

This work has been partially supported by European Community through the FET-Proactive Project {\it MoQuaS} (contract N. 610449), by the Italian Ministry for Research (MIUR) through the FIR grant RBFR13YKWX, and by the Alexander von Humboldt Foundation. The authors acknowledge useful discussions with Andrea Candini.


\end{document}